\documentclass[a4paper,11pt]{article}
\pdfoutput=1 

\usepackage{jheppub} 
                     
\usepackage[T1]{fontenc} 
\usepackage[dvipsnames]{xcolor}
\usepackage{subfigure}
\usepackage{braket}
\usepackage{bbm}
\usepackage[normalem]{ulem}
\usepackage{hyperref}

\def\Tr{\mathop{\mbox{\normalfont Tr}}\nolimits}

\begin{document}

\title{\boldmath Symmetry resolution of the computable cross-norm negativity of two disjoint intervals in the massless Dirac field theory
}

\vspace{.5cm}

\author{Andrea Bruno$^{1}$, Filiberto Ares$^{2}$, Sara Murciano$^{3,4}$, Pasquale Calabrese$^{2,5}$}
\affiliation{$^{1}$Scuola Normale Superiore and Dipartimento di Fisica dell'Universit\`a di Pisa, I-56127 Pisa, Italy}
\affiliation{$^2$SISSA and INFN Sezione di Trieste, via Bonomea 265, 34136 Trieste, Italy.}
\affiliation{$^{3}$Walter Burke Institute for Theoretical Physics, Caltech, Pasadena, CA 91125, USA}
\affiliation{$^{4}$Department of Physics and IQIM, Caltech, Pasadena, CA 91125, USA}
\affiliation{$^{5}$International Centre for Theoretical Physics (ICTP), Strada Costiera 11, 34151 Trieste, Italy.}

\abstract{
We investigate how entanglement in the mixed state of a quantum field theory can be described using the cross-computable norm or realignment (CCNR) criterion, employing a recently introduced  negativity.  
We study its symmetry resolution for two disjoint intervals in the ground state of the massless Dirac fermion field theory, extending previous results for the case of adjacent intervals. By applying the replica trick, this problem boils down to computing the charged moments of the realignment matrix. We show that, for two disjoint intervals, they correspond to the partition function of the theory on a torus with a non-contractible charged loop. 
This confers a great advantage compared to the negativity based on the partial transposition, for which the Riemann surfaces generated by the replica trick have higher genus. This result empowers us to carry out the replica limit, yielding  analytic expressions for the symmetry-resolved CCNR negativity. 
Furthermore, these expressions provide also  the symmetry decomposition of other related quantities such as the operator entanglement of the reduced density matrix or the reflected entropy. }
\maketitle
\section{Introduction}

While the entanglement entropy is the canonical quantity to measure 
the entanglement between two parts that share a pure state~\cite{afov-08, ccd-09, ecp-10, laflorencie-14}, it does
not longer serve as an entanglement quantifier when they are in a 
mixed state, since it also detects the classical correlations present 
in this kind of states. Opposite to the pure state case, there is no
a canonical measure of bipartite entanglement in mixed states. By relying on different criteria, several quantities have been proposed
to discern the separability of a mixed state and measure its correlations. Among them, we can mention the PPT negativity~\cite{vw-02, plenio-05}, which
employs the  Positive Partial Transposition (PPT) criterion~\cite{peres-96, hhh-96, simon-00}, and the 
CCNR negativity, which is based on the Computable Cross-Norm or Realignment (CCNR) criterion~\cite{rudolph-03, cw-03}.

The PPT negativity has been largely studied in extended quantum systems: 
free-fermions~\cite{ez-15, ez-16, ssr-17, ssr-17-2, srrc-19, sr-19, sr-19-2}, chains of harmonic oscillators~\cite{aepw-02, fcgsa-08, cfgsa-08, mrpr-09, ez-14, dnct-16}, and spin chains~\cite{wmpv-09, bbs-10, bbsj-12, rac-16-2, mac-17, lg-19, trc-20, waa-20, slkfv-21, mvdc-22}, or $(1+1)$ integrable and conformal field
theories (CFTs)~\cite{cct-12, cct-13, ctt-13, alba-13, cct-15, rac-16, ctc-16-2, asv-21, rockwood-22,rmtc-23, rmc-23, bfcad-16, cadfds-19} to cite some of them. 
On the other hand, the CCNR negativity has received little attention in this area and only very recent works have focused on it, such as in CFTs~\cite{yl-23}, holography~\cite{mrw-22}, free fermionic and bosonic chains~\cite{bp-23}, and topological phases~\cite{yl-23-2}. The CCNR negativity is intimately connected with other information based quantities such as operator
entanglement~\cite{pp-07,Pizorn2009,zanardi2000entangling,zanardi2001entangling,dubail}, which characterises the complexity of operators like the reduced density matrix, or the reflected entropy, which was introduced to study mixed state correlations in holographic theories~\cite{df-21} and it is useful to investigate tripartite entanglement~\cite{ar-20, hps-21}.

Among the mixed states, a remarkable and widely used example consists of two disjoint spatial parts of an extended quantum system in a pure state. In fact, the reduced density matrix of the two subsystems is generally mixed, and
one must resort to the PPT or the CCNR negativities to probe the entanglement between them. This question has been particularly investigated to a great extent in the ground state of $(1+1)$ CFTs. One of the most remarkable properties of entanglement entropy and negativities is that they encapsulate the universal data of the CFT; it is well-known that the 
entanglement entropy of one interval is proportional to the central charge~\cite{hlw-94, cc-04}. In the case of the PPT and CCNR negativities, they encode the
full operator content of the CFT~\cite{cct-12, cct-13, yl-23}, as it is clear when  we apply the replica trick~\cite{cc-04} to calculate them. For the PPT negativity, this method boils down to
determining the moments of the partial transpose of the reduced density matrix of the two intervals, which are identified with the partition function of the field theory on a family of complicated higher genus Riemann surfaces~\cite{cct-12, cct-13}. The PPT negativity can be obtained by analytically continuing these partition functions in the genus and then taking a specific limit. This is in general a difficult problem and the analytic continuation to get the PPT negativity is still an open issue. This calculation is, however, easier for the CCNR negativity. By applying the replica trick, this quantity can be derived from the moments of the two-interval realignment matrix, which can be cast as the partition
function of the theory on a torus whose modulus is re-scaled by the 
number of copies considered~\cite{yl-23}. In this case, the replica limit to get 
the CCNR negativity is straightforward. 

A recent noteworthy development is how entanglement is distributed between
the symmetry sectors of a theory that presents a global internal symmetry. This question, initially posed in Refs.~\cite{lr-14, goldstein, xavier}, has been examined from many different perspectives~\cite{riccarda,kusf-20,ms-20,trac-20,mdgc-20,bpc-21,capizzi,dhc-20,znm-20,regnault}, especially in CFTs \cite{Belin-Myers-13-HolChargedEnt,nonabelian-21,crc-20,Chen-21,Capizzi-Cal-21,boncal-21,eim-d-21,mt-22,acgm-22,Ghasemi-22,fac-23,gmnsz-23,northe-23,kmop-23,cmc-23}. The main motivation behind those works is that the symmetry resolution of entanglement provides a better understanding of some properties of 
many-body quantum systems, as it has also been testified experimentally~\cite{neven,fis, azses, vitale, rath-23}. After the partial transpose operation, the reduced density matrix can be decomposed into the contributions of the charge imbalance between the two subsystems. This was first proven and applied to obtain the symmetry-resolved PPT negativity in CFTs with a global $U(1)$ charge in Ref.~\cite{cgs-18} and then extended to different systems and situations~\cite{mbc-21, fg-19, pbc-22, chen-22, chen-22-2, mhms-22, foligno-23,gy-23}, also at experimental level~\cite{neven}. 
So far, the symmetry resolution of the CCNR negativity has only been considered in Ref.~\cite{bp-23} for the ground state of free fermionic and bosonic theories when the two intervals share a common endpoint. 
In this context, it has shown that 
the realignment matrix can be decomposed in the sectors of the super-charge operator associated to the symmetry resolution of the operator entanglement of the reduced density matrix~\cite{rath-23, mdc-23}. 

The goal of the present work is to generalize the results of Ref.~\cite{bp-23} to disjoint subsystems in the ground state of the massless free Dirac fermion. The symmetry-resolved CCNR negativity can be computed from the charged moments of the realignment matrix. 
As we will show, when the intervals are disjoint, the latter correspond to the partition function on a torus with modulus re-scaled by the number of copies and a charged loop inserted along one of its non-contractible cycles. Thus the calculation of the charged moments of the realignment matrix is reduced to a (non-trivial) textbook problem.
From this result, we obtain analytic expressions for the symmetry resolution of the two disjoint interval CCNR negativity, which also
directly gives the symmetry decomposition of other interesting quantities such as the operator entanglement or the reflected entropy~\cite{reflected2,chhjw-21,bc-20}, that have not been previously considered.

The paper is organised as follows. In Sec.~\ref{sec:ccnr_neg}, we introduce the basic definitions of the realignment matrix and the CCNR negativity, we discuss its relation with operator entanglement and we explain how to calculate it in CFT. We also illustrate how to resolve the CCNR negativity in the presence of a global $U(1)$ charge and the general approach to compute it in terms of the charged moments of the realignment matrix. We review the results of Ref.~\cite{bp-23} for the case of adjacent intervals. In Sec.~\ref{sec:new}, we consider the ground state of the massless Dirac fermion, and we analytically calculate the charged moments of the realignment matrix for two disjoint intervals, by identifying them with the torus partition function of the theory with a charged loop. We then use this result to derive the symmetry-resolved CCNR negativity. We check the analytic expressions that we obtain with exact numerical calculations in the tight-binding model, a lattice realisation of the massless Dirac theory. We conclude in Sec.~\ref{sec:conclusions} with some remarks and future prospects. We also include two appendices where we give the detailed derivation of some identities that we employ in the main text and describe the numerical method that we apply to check the analytic results.

\begin{figure}[t!]
    \centering
    \includegraphics[width=0.8\textwidth]{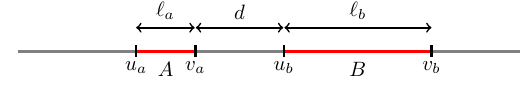}
    \caption{Schematic representation of the system that we study. We consider a one dimensional critical system described by a CFT and we analyse the entanglement between two intervals $A=[u_a, v_a]$ and $B=[u_b, v_b]$ of lengths $\ell_a=|u_a-v_a|$ and $\ell_b=|u_b-v_b|$ separated by a distance $d=|v_a-u_b|$.} 
    \label{fig:system}
\end{figure}

\section{Computable cross-norm (CCNR) negativity}\label{sec:ccnr_neg}

In this section, we define the concept of CCNR negativity and we review the main results about this quantity in the ground state of $(1+1)$ CFTs, both for the total~\cite{yl-23} and the symmetry-resolved version~\cite{bp-23}.

\subsection{Definition and previous results for CFTs from Ref.~\cite{yl-23}}\label{sub:total}
\paragraph{Definitions:} Let us take a one-dimensional system in a state described by the density matrix $\rho$. As we show in Fig.~\ref{fig:system}, we will consider two disjoint intervals, $A$ and $B$, whose Hilbert spaces are $\mathcal{H}_A$ and $\mathcal{H}_B$, respectively. We denote with $\rho_{AB}$ the reduced density matrix of the subsystem $A\cup B$ obtained from the trace over the complementary set $C$,
\begin{equation}
    \rho_{AB}=\mathrm{Tr}_{C}\rho.
\end{equation}
If $\{\ket{\text{a}}\}$ and $\{\ket{\text{b}}\}$ are bases for $\mathcal{H}_A$ and $\mathcal{H}_B$ respectively, then we can define the realignment matrix $R$ as the linear operator $R:\mathcal{H}_B\otimes\mathcal{H}_B\rightarrow\mathcal{H}_A \otimes \mathcal{H}_A$ with entries
\begin{equation}\label{eq:def_realignment}
\bra{\text{a}}\bra{\text{a}'}R\ket{\text{b}}\ket{\text{b}'}\equiv\bra{\text{a}}\bra{\text{b}}\rho_{AB}\ket{\text{a}'}\ket{\text{b}'}.
\end{equation}
In a generic finite-dimensional system, $R$ is not necessarily a square matrix, unless the dimensions of the Hilbert spaces $\mathcal{H}_A$ and $\mathcal{H}_B$ are the same, i.e. when $A$ and $B$ have the same size. It can be proven that, if $\rho_{AB}$ is separable, then~\cite{rudolph-03, cw-03}
\begin{equation}\label{eq:ccnr_def}
    \lVert R \rVert \equiv \text{Tr}[\sqrt{RR^{\dagger}}]\leq 1,
\end{equation}
where $\lVert R \rVert $ denotes the trace norm.
Hence a necessary (but not suffcient) condition for the state shared by $A$ and $B$ to be entangled is 
$\lVert R \rVert >1$,  known as CCNR criterion~\cite{rudolph-03, cw-03}. This criterion allows to define the following quantity
\begin{equation}\label{eq:ccnr_neg_def}
\mathcal{R}=\max(\mathcal{E}, 0),\quad \mathcal{E}=\log \lVert R \rVert,
\end{equation}
usually dubbed CCNR negativity, as an entanglement probe in a mixed state: if $\mathcal{R}>0$, then $\rho_{AB}$ is entangled, and $\mathcal{R}=0$ if $\rho_{AB}$ is separable.
However,
if $\mathcal{R}=0$ it is not guaranteed that the state is separable.

The trace norm $\lVert R \rVert$ can be obtained by analytically continuing the moments of $RR^\dagger$, $\mathrm{Tr}[(RR^{\dagger})^n]$, and then taking the limit $n\to 1/2$. Mimicking the usual approach to study the negativity based on the PPT criterion~\cite{cct-12, cct-13}, one can introduce the CCNR R\'enyi negativities~\cite{yl-23}
\begin{equation}\label{eq:ccnr_renyi}
    \mathcal{E}_n=\log\mathrm{Tr}[(RR^{\dagger})^n].
\end{equation}
Taking into account Eq.~\eqref{eq:ccnr_neg_def}, the CCNR negativity can be then calculated from 
\begin{equation}
    \mathcal{E}=\lim_{n\to 1/2}\mathcal{E}_n.
    \label{csi}
\end{equation}

Beyond the conceptual similarity with the PPT negativity, we can also relate $\mathcal{E}_n$ to other information theoretic quantities such as the operator entanglement~\cite{mrw-22, bp-23}. Indeed, the reduced density matrix $\rho_{AB}$ can be vectorised using the Choi-Jamiolkowski isomorphism~\cite{choi, jamiolkowski},
\begin{equation}\label{eq:vect}
\ket{\rho_{AB}}=\frac{1}{\sqrt{\Tr(\rho_{AB}^2)}}\sum_{{\rm a},{\rm b}, {\rm a'}, {\rm b'}} \bra{\rm ab} \rho_{AB} \ket{\rm a'b'} \ket{\rm ab}\ket{\rm a'b'},
\end{equation}
by doubling the original Hilbert space $\mathcal{H}_A\otimes \mathcal{H}_B$. Note that $\ket{\rho_{AB}}$ has norm one. From it,
we can define the super-reduced density matrix $\sigma_{AA'}$ as
\begin{equation}\label{eq:super}
\sigma_{AA'}=\Tr_{BB'}(\ket{\rho_{AB}}\bra{\rho_{AB}}),
\end{equation}
where $\Tr_{BB'}$ denotes the partial trace over $\mathcal{H}_B$ and its replica. Combining the definitions of the realignment matrix $R$ and of the super-reduced density matrix $\sigma_{AA'}$, it is straightforward to show that (see Appendix~\ref{app:proofs} for more details) 
\begin{equation}\label{eq:ReqOE}
RR^\dagger = \Tr(\rho_{AB}^2)\sigma_{AA'}.
\end{equation}
If we plug this identity in the definition~\eqref{eq:ccnr_renyi} of the CCNR R\'enyi negativities, then we find that
\begin{equation}\label{eq:SOEtoE}
\mathcal{E}_n=(1-n)S_n^{\rm OE}(\rho_{AB})+n\log \Tr(\rho_{AB}^2),
\end{equation}
where $S_n^{\rm OE}(\rho_{AB})$ is the R\'enyi operator entanglement of $\rho_{AB}$,
that is the R\'enyi entropy of the super-reduced density matrix $\sigma_{AA'}$,
\begin{equation}
S_n^{\rm OE}(\rho_{AB})=\frac{1}{1-n}\log\Tr(\sigma_{AA'}^n).
\end{equation}
Eq.~\eqref{eq:SOEtoE} shows the direct connection between the CCNR negativity and the operator entanglement of the reduced density matrix $\rho_{AB}$. Refs.~\cite{mrw-22} and~\cite{bp-23} discuss in detail how both quantities also intersect another tool to study quantum correlations in mixed states, the reflected entropies~\cite{df-21}, which were mainly introduced in the holographic context.

\begin{figure}[t!]
    \centering
    \includegraphics[width=5.5cm]{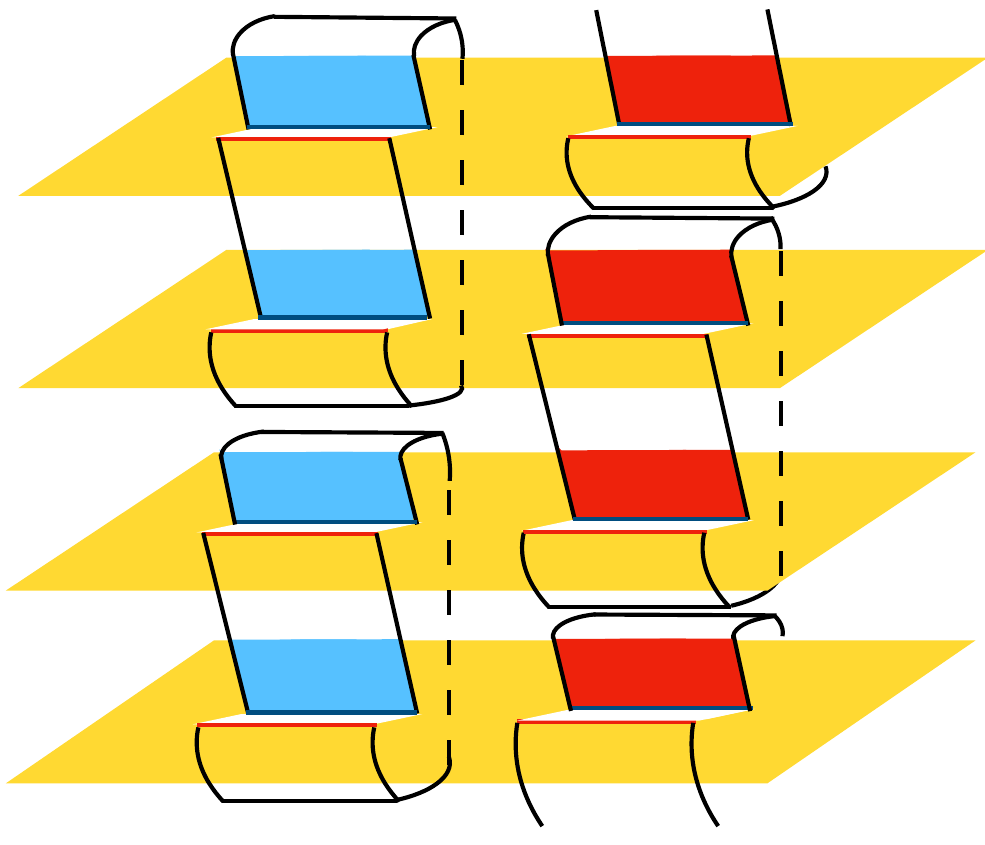}
    \caption{Pictorial representation of the Riemann surface $\mathcal{M}_{2n}$ for $n=2$, whose partition function provides $Z_2=\Tr(RR^\dagger)^2$. The cut along the left (right) interval of each sheet is identified with the corresponding one of the lower (upper) sheet. }
    \label{fig:riemann1}
\end{figure}

\paragraph{CFT:} Let us now review how the CCNR R\'enyi negativities defined in Eq.~\eqref{eq:ccnr_renyi} can be computed in the ground state of a $(1+1)$ CFT with central charge $c$. As shown in Ref.~\cite{yl-23} using the path integral representation of $\rho_{AB}$~\cite{cc-04}, the object $Z_n=\mathrm{Tr}[(RR^{\dagger})^n]$ is, for positive integer $n$, the partition function of the theory on the $2n$-sheeted Riemann surface $\mathcal{M}_{2n}$ in which the cut along the interval $A$ ($B$) of each sheet is sewed with the corresponding one of the lower (upper) sheet as we depict in Fig.~\ref{fig:riemann1} for the case $n=2$. The surface $\mathcal{M}_{2n}$ has genus one for any integer value of $n$. This is a great advantage with respect to the PPT negativity, in which the Riemann surface that arises when 
applying the replica trick has a genus that increases with the 
index $n$~\cite{cct-12, cct-13}. As discussed in Ref.~\cite{yl-23}, if we take $2n$ replicas of the original theory, then the partition function on the surface $\mathcal{M}_{2n}$ can be written as the correlation function of the twist fields $\mathcal{T}_{2n}'(z)$ and $\widetilde{\mathcal{T}}_{2n}'(z)$ inserted at the end-points of $A=[u_a,v_a]$ and $B=[u_b,v_b]$ respectively,
\begin{equation}\label{eq:Zn_R}
    Z_n=\langle \mathcal{T}'_{2n}(u_a) \mathcal{T}'_{2n}(v_a) \widetilde{\mathcal{T}}_{2n}'(u_b) \widetilde{\mathcal{T}}_{2n}'(v_b) \rangle.
\end{equation}
We notice that $\mathcal{T}'_{2n}(z)$ and $\widetilde{\mathcal{T}}'_{2n}(z)$ are different from the twist fields that implement the cyclic permutations between replicas in the standard computation of the R\'enyi entanglement entropy~\cite{cc-04} and the PPT negativities~\cite{cct-12, cct-13}, usually denoted as $\mathcal{T}_{n}(z)$ and $\widetilde{\mathcal{T}}_{n}(z)$, and, therefore, we use the prime notation to refer to them. According to the way the sheets of the Riemann surface $\mathcal{M}_{2n}$ are glued, see Fig.~\ref{fig:riemann1}, the twist fields $\mathcal{T}_{2n}'(z)$ implement the following permutations among the copies
\begin{align*}
    \mathcal{T}_{2n}':(1 &\leftrightarrow 2, 3 \leftrightarrow 4, \dots, 2n-1 \leftrightarrow 2n),
\end{align*}
while, for the fields $\widetilde{\mathcal{T}}_{2n}'(z)$, 
\begin{align*}    
    \widetilde{\mathcal{T}}_{2n}':(2 \leftrightarrow 3, 4 \leftrightarrow 5 \dots, 2n \leftrightarrow 1).
\end{align*}
Both $\mathcal{T}'_{2n}(z)$ and $\widetilde{\mathcal{T}}'_{2n}(z)$ are spinless primaries and their conformal dimension is given by~\cite{yl-23}
\begin{equation}\label{eq:conf_dim_ccnr_twist}
h_{\mathcal{T}'_{2n}}=nh_{\mathcal{T}_{2}}=\frac{nc}{16},
\end{equation}
where $h_{\mathcal{T}_{n}}$ is the conformal dimension of the standard
twist fields $\mathcal{T}_{2n}(z)$ and $\widetilde{\mathcal{T}}_{2n}(z)$~\cite{cc-04}, 
\begin{equation}\label{eq:conf_dim_standard_twist}
 h_{\mathcal{T}_{n}}=\frac{c}{24}\left(n-\frac{1}{n}\right).
\end{equation}

To simplify the computation of the correlator~\eqref{eq:Zn_R}, it is helpful to perform a global conformal transformation that maps the end-points $(u_a, v_a, u_b, v_b)$ to $ (0, x, 1, \infty)$, where $x$ is the cross-ratio defined as
\begin{equation}\label{crossratio}
    x=\frac{(v_a-u_a)(v_b-u_b)}{(v_b-v_a)(u_b-u_a)}.
\end{equation}
Taking into account that $\mathcal{T}_{2n}'(z)$ and $\widetilde{\mathcal{T}}_{2n}'(z)$ are primaries of dimension \eqref{eq:conf_dim_ccnr_twist}, Eq.~\eqref{eq:Zn_R} can be rewritten as
\begin{equation}\label{eq:referencexprefactor}
    Z_n=\bigg( \frac{x}{(v_a-u_a)(v_b-u_b)} \bigg)^{\frac{nc}{4}} \langle \mathcal{T}'_{2n}(0) \mathcal{T}'_{2n}(x) \widetilde{\mathcal{T}}'_{2n}(1) \widetilde{\mathcal{T}}'_{2n}(\infty) \rangle.
\end{equation}
The twist-field correlator on the right hand side of this expression corresponds to the partition function on the surface $\mathcal{M}_{2n}$ with branch points at $(0,x,1,\infty)$. We start our analysis by taking $n=1$. In this case, we can map the surface $\mathcal{M}_2$  to a flat torus with modular parameter $\tau$ related to $x$ by
\begin{equation}
    \tau=i\frac{K(1-x)}{K(x)},
    \label{taulaw}
\end{equation}
where $K(x)$ is the complete elliptic integral of the first kind,
through the conformal transformation
\begin{equation}\label{map}
    \omega(t)=\frac{\wp(t)-\wp((1+\tau)/2)}{\wp(1/2)-\wp((1+\tau)/2)},
\end{equation}
where $\wp(t)$ is the Weierstrass elliptic function. Eq.~\eqref{map} establishes a one-to-two correspondence 
between points on the flat torus, $t\in\mathbb{C}/(\mathbb{Z}+\tau \mathbb{Z})$, and points on the surface $\mathcal{M}_{2}$, $\omega(t)\in \mathcal{M}_{2}$. Applying the transformation~\eqref{map}, the four-point correlation function appearing in Eq.~\eqref{eq:referencexprefactor} can be written as~\cite{lm-01}
\begin{equation}\label{eq:partR2_Ztau}
   \langle \mathcal{T}'_{2}(0) \mathcal{T}'_{2}(x) \widetilde{\mathcal{T}}'_{2}(1) \widetilde{\mathcal{T}}'_{2}(\infty) \rangle = c_1Z(\tau) |x(1-x)|^{-c/12},
\end{equation}
where $Z(\tau)$ is the partition function on the flat torus and $c_1$ is a non-universal function depending on the underlying lattice discretisation of the conformal field theory. The factor $|x(1-x)|^{-c/12}$ is the conformal anomaly and takes into account the conical singularities of the surface $\mathcal{M}_2$ at the branch points~\cite{lm-01, msz-17}. Plugging Eq.~\eqref{eq:partR2_Ztau} into Eq.~\eqref{eq:referencexprefactor}, we find 
\begin{equation}\label{eq:Ztau}
    Z_1=\frac{c_1Z(\tau)}{[\ell_b\ell_a(u_b-u_a)(v_b-v_a)(v_b-u_a)(u_b-v_a)]^{c/12}}.
\end{equation}

Eq.~\eqref{eq:Ztau} can be straighforwardly generalised to any value of $n$. In fact, using the path integral formalism, the crucial finding of Ref.~\cite{yl-23} is that the Riemann surface $\mathcal{M}_{2n}$ is topologically equivalent to a torus of period $n\tau$ for any positive integer value of $n$. Therefore, the partition function $Z_n$ of the surface $\mathcal{M}_{2n}$ can be directly obtained from the expression~\eqref{eq:Ztau} for $n=1$ by replacing the modulus $\tau\mapsto n\tau$ and the central charge  $c\mapsto nc$,
\begin{equation}\label{eq:Zntau}
    Z_n=\frac{c_1^nZ(n\tau)}{[\ell_b\ell_a(u_b-u_a)(v_b-v_a)(v_b-u_a)(u_b-v_a)]^{nc/12}}.
\end{equation}
This also shows why computing the CCNR negativity is easier than the PPT negativity, since in the latter case we need to evaluate a partition function on a $(n-1)$-genus Riemann surface, while the former only involves computations on a toroidal surface. Moreover, Eq.~\eqref{eq:Zntau} can be analytically continued to real values of $n$ to obtain the CCNR negativity~\eqref{eq:ccnr_neg_def} taking the limit $n\to 1/2$. However, there is not free lunch here: if $A$ and $B$ are disjoint intervals, $Z_n<1$ for several values of $n$, also in the limit $n\to 1/2$. Therefore, even though it is easier to compute with respect to the PPT negativity, the CCNR negativity often fails to establish whether the state $\rho_{AB}$ is entangled or not. 

It is interesting to report explicitly the result in the limit $v_a\to u_b$, in which the intervals $A$ and $B$ are adjacent.  
In this case, the partition function in Eq.~\eqref{eq:Zn_R} reduces to the three-point correlation function~\cite{yl-23}
\begin{equation}\label{eq:corr_function_adjacent}
    Z_n=\langle \mathcal{T}'_{2n}(u_a) \mathcal{T}_{n}^{\otimes 2}(v_a) \widetilde{\mathcal{T}}'_{2n}(v_b)  \rangle,
\end{equation}
whose expression, up to a non-universal constant, is fixed by the conformal dimension of $\mathcal{T}'_{2n}(z)$, $\widetilde{\mathcal{T}}_{2n}'(z)$ $\mathcal{T}_{n}^{\otimes 2}(z)$, see Eqs.~\eqref{eq:conf_dim_ccnr_twist} and~\eqref{eq:conf_dim_standard_twist} respectively, and it reads
\begin{equation}
    Z_n \propto (\ell_a\ell_b)^{-\frac{c}{6}(n-\frac{1}{n})}(\ell_a+\ell_b)^{-\frac{c}{12}\left(n+\frac{2}{n}\right)}.
    \label{joint neutral}
\end{equation}
Taking the replica limit $n\to 1/2$, we obtain that $\mathcal{R}$ is always a positive quantity, indicating that the state $\rho_{AB}$ is entangled when $A$ and $B$ are adjacent.

\subsection{Symmetry-resolved CCNR negativity: Definition and previous results from Ref.~\cite{bp-23}}\label{subsec:SR}

\paragraph{Definitions:}

When the system displays a global $U(1)$ symmetry, the CCNR negativity associated to $\rho_{AB}$ can be split into the contribution of individual charge sectors; 
this  decomposition mirrors the approach taken for the symmetry resolution of the entanglement entropy~\cite{lr-14, goldstein, xavier} or the PPT negativity~\cite{cgs-18}. 
Let the charge $Q$ be the generator of the $U(1)$ symmetry in the total system and assume that it is local, in the sense that it is the sum of the charges within the subregions, i.e.
$Q=Q_A+Q_B+Q_C$. When $\rho_{AB}$ results from tracing out the subsystem $C$ and the state of the total system is an eigenstate of $Q$, we have
\begin{equation}
    [Q_{A}+Q_B, \rho_{AB}]=0.
    \label{qqq}
\end{equation}
To define the symmetry-resolved CCNR negativity, we should analyse how
the realignment matrix $R$ decomposes in the symmetry sectors. As we already stressed, this is in general not a square matrix and it is difficult to work with it. We can instead consider the product $RR^\dagger$, which is a square matrix and admits a block diagonal decomposition in the eigenbasis of the superoperator~\cite{mrw-22, bp-23}
\begin{equation}\label{eq:superQ}
 \mathcal{Q}_{A}=Q_A\otimes \mathbb{I}-\mathbb{I}\otimes Q_{A}.
\end{equation} 
This is not surprising since, according to Eq.~\eqref{eq:ReqOE}, $RR^\dagger$ is
equal to $\sigma_{AA'}$ up to a normalisation factor and, as shown in Ref.~\cite{rath-23}, the super-reduced density matrix in the presence of a global $U(1)$ symmetry takes the form
\begin{equation}
  \sigma_{AA'}=\bigoplus_q p(q)\sigma_{AA'}(q),\qquad  \sigma_{AA'}(q)= \frac{\Pi_q \sigma_{AA'}\Pi_q}{p(q)}.
    \label{sigmasplit}
\end{equation}
Here $\Pi_q$ is the projector onto the sector of charge $q$ of $\mathcal{Q}_A$ and $p(q)=\text{Tr}_{AA'}[\Pi_q \sigma_{AA'} ]$ normalises $\sigma_{AA'}$ such that $\Tr(\sigma_{AA'}(q))=1$. Since
$RR^\dagger$ does not have unit trace, the symmetry resolution of the
CCNR negativity is ambiguous. A natural choice is to write $RR^\dagger$
as 
\begin{equation}\label{eq:symm_dec_RRdag}
RR^\dagger =\bigoplus_q p(q) \Sigma(q),\quad \Sigma(q)=\frac{\Pi_q RR^\dagger \Pi_q}{p(q)}.
\end{equation}
We then define the {\it symmetry-resolved CCNR R\'enyi negativity} as~\cite{bp-23}
\begin{equation}\label{eq:def_SRCCNR}
\mathcal{E}_n(q)=\log\Tr(\Sigma(q)^n).
\end{equation}
Combining Eqs.~\eqref{sigmasplit} and~\eqref{eq:symm_dec_RRdag} with~\eqref{eq:ReqOE}, we have $\Sigma(q)=\Tr(\rho_{AB}^2)\sigma_{AA'}(q)$. If we insert it in the definition
of Eq.~\eqref{eq:def_SRCCNR}, then we find
\begin{equation}
    \mathcal{E}_n(q)\equiv (1-n) S_n^{\text{OE}}(q)+n \log \mathrm{Tr}(\rho_{AB}^2),
    \label{eq:realrenyi}
\end{equation}
where $S_n^{\rm OE}(q)$ is the symmetry-resolved operator entanglement~\cite{rath-23}
\begin{equation}\label{eq:SRCCNR_SROE}
    S_n^{\text{OE}}(q)\equiv \frac{1}{1-n}\log\text{Tr}[\sigma_{AA'}(q)^n].
\end{equation}
Eq.~\eqref{eq:SRCCNR_SROE} is analogue to the identity in Eq.~\eqref{eq:SOEtoE} between the total operator entanglement and the CCNR R\'enyi negativity.

The goal of this paper is to study $\mathcal{E}_n(q)$ in a field theory setup. In that case, its direct calculation from the definition of Eq.~\eqref{eq:def_SRCCNR} is a hard task, which can be bypassed by applying the approach used in Ref.~\cite{goldstein} to obtain the symmetry resolution of the entanglement entropies. This is based on the Fourier representation of the projection operator $\Pi_q$. If we perform it, we then have 
\begin{equation}
    \mathcal{Z}_n(q)\equiv \mathrm{Tr}[\Pi_q (RR^\dagger)^n]=\int_{-\pi}^{\pi} \frac{d\alpha}{2\pi}e^{-iq\alpha}Z_{n}(\alpha), \qquad Z_{n}(\alpha)=\text{Tr}[(RR^{\dagger})^n e^{i\alpha\mathcal{Q}_A}],
    \label{eq:ft}
\end{equation}
i.e. we can relate the moments of the projected partition function $ \mathcal{Z}_n(q)$ to the charged moments $Z_n(\alpha)$ of $RR^{\dagger}$. Therefore, Eq.~\eqref{eq:def_SRCCNR} can be rewritten as 
\begin{equation}\label{eq:SRCCNR_fourier}
     \mathcal{E}_n(q)=\log \frac{\mathcal{Z}_n(q)}{\mathcal{Z}^n_1(q)}+n\log\mathrm{Tr}(\rho_{AB}^2).
\end{equation}
The main focus of the following sections is to first review the results about $Z_{n}(\alpha)$ and $\mathcal{E}_n(q)$ found in Ref.~\cite{bp-23} for two adjacent intervals in a free fermionic theory and, later, we will show how this quantity can be computed in the disjoint geometry.

\paragraph{CFT (Adjacent case):} 
As we have just seen, the symmetry-resolved CCNR R\'enyi negativities
~\eqref{eq:def_SRCCNR} can be evaluated from the charged moments $Z_n(\alpha)$ of the matrix
$RR^\dagger$. The main advantage of this approach is that, in field 
theory, the operator $e^{i\alpha \mathcal{Q}_A}$ can be easily implemented within the geometrical framework that we have introduced before to calculate
the neutral moments $Z_n(0)=Z_n$. The charged moments $Z_n(\alpha)$ have been studied in 
Ref.~\cite{bp-23} for free fermionic and bosonic theories with a
$U(1)$ symmetry when $A$ and $B$ are adjacent intervals. We report here the result for the fermionic case:  $Z_n(\alpha)$ is given by a modification of the twist-field correlation function~\eqref{eq:corr_function_adjacent},
\begin{equation}\label{eq:vertex}
Z_n(\alpha)=\langle \mathcal{T}_{2n}'(u_a) (\mathcal{T}_{n, \alpha}\otimes \mathcal{T}_{n,-\alpha})(v_a) \widetilde{\mathcal{T}}_{2n}'(v_b)\rangle,
\end{equation}
where $\mathcal{T}_{n,\alpha}(z)$ is a primary field with conformal dimension
\begin{equation}\label{eq:dim_composite_fields}
 h_{n,\alpha}=h_n+\frac{\alpha^2}{8\pi^2n},
\end{equation}
that takes into account the effect of the operator $e^{i\alpha\mathcal{Q}_A}$.
%
Using the well-known expression for the correlation function of three
primary fields, one obtains
\begin{equation}\label{eq:gilles}
    Z_n(\alpha)\propto \left(\frac{\ell_a\ell_b}{\ell_a+\ell_b}\right)^{-\frac{\alpha^2}{2\pi^2n}}\left(\ell_a\ell_b\right)^{-\frac{1}{6}(n-\frac{1}{n})}(\ell_a+\ell_b)^{-\frac{1}{12}\left(n+\frac{2}{n}\right)}.
\end{equation}
By performing the Fourier transform in Eq.~\eqref{eq:ft}, the symmetry-resolved CCNR negativity then reads
\begin{equation}\label{eq:gillesadj}
    \mathcal{E}_n(q)=\mathcal{E}_n-\frac{1-n}{2}\log \log \frac{\ell_a\ell_b}{\ell_a+\ell_b}+\frac{1-n}{2}\log \frac{\pi n^{1/(1-n)}}{2}+O\left(\log^{-2}\frac{\ell_a\ell_b}{\ell_a+\ell_b}\right).
\end{equation}
The fact that, at leading order, the
symmetry-resolved CCNR negativity does not depend on the charge sector is reminiscent of what happens for
the symmetry-resolved entanglement entropies and is known as equipartition of
entanglement~\cite{xavier}. A similar behaviour has been found also for the symmetry resolution of the PPT negativity in terms of the charge imbalance
between two adjacent subsystems \cite{mbc-21,foligno-23}.

\section{Symmetry resolution of CCNR negativity in disjoint intervals}\label{sec:new}
After this introduction about the concept of CCNR negativity and its decomposition in the presence of a global $U(1)$ symmetry, we now specialise to its computation for two disjoint intervals in the ground state of the free massless Dirac field theory. Its action is 
\begin{equation}
\label{eq:lagrangianDirac}
\mathcal{S}=\int {\rm d}x_0 {\rm d}x_1 \bar{\psi}\gamma^\mu \partial_\mu \psi ,
\end{equation}  
where $\bar{\psi}=\psi^\dagger \gamma^0$. The $\gamma^\mu$ matrices can be represented 
in terms of the Pauli matrices as $\gamma^0=\sigma_1$ and $\gamma^1=\sigma_2$.
The invariance of the action of Eq.~(\ref{eq:lagrangianDirac}) under $\psi\mapsto e^{i \alpha}\psi$ 
and $\bar{\psi}\mapsto e^{-i \alpha}\bar{\psi}$ corresponds to the conservation of the $U(1)$ charge $Q=\int {\rm d}x_1 \psi^\dagger\psi $.  We will first determine the charge moments $Z_n(\alpha)$ of $RR^\dagger$ corresponding to this symmetry and, from them, we will compute the symmetry-resolved CCNR negativity.

We remark here that, when applying the replica trick to a theory 
containing fermionic fields, one needs to be careful about the 
boundary conditions~\cite{hlr-12, ctc-16}. If we are interested in the modular 
invariant theory, to obtain the neutral moments $Z_n(0)$ one should sum over the partition functions corresponding to all possible choices of boundary conditions, Neveu-Schwarz (NS) and Ramond (R), around the non-contractible cycles of the Riemann surface $\mathcal{M}_{2n}$. 
However, we can also focus on computing the partition function with 
a specific set of boundary conditions around the cycles. In the following, we use this second approach and we impose NS boundary conditions on both cycles of $\mathcal{M}_{2n}$. In this case, a microscopic lattice realisation corresponds to the tight-binding model. Using the latter, we can also cross-check our field theory analytical predictions against exact numerical computations. 

The partition function $Z(\tau)$ of the 
theory~\eqref{eq:lagrangianDirac} on the flat torus of modulus $\tau$ and with NS boundary conditions can be obtained from 
\begin{equation}\label{eq:NS_flat_torus_part_func}
Z(\tau)=\left|\Tr_{\mathcal{H}_{\rm NS}}(q^{L_0-\frac{c}{24}})\right|^2, \quad q=e^{2\pi i \tau},
\end{equation}
where $\mathcal{H}_{\rm NS}$ is the Hilbert space of the NS sector
and $L_0$ is the zero mode of the Virasoro generators on the complex plane. Eq.~\eqref{eq:NS_flat_torus_part_func} gives~\cite{blumenhagen}
\begin{equation}
Z(\tau)=\left|\frac{\theta_3(0|\tau)}{\eta(\tau)}\right|^2,
\end{equation}
where $\theta_\nu(z| \tau)$ and $\eta(\tau)$ are the Jacobi theta and the Dedekind eta functions, respectively.
Plugging this result in Eq.~\eqref{eq:Zntau}, we obtain the explicit expression of the neutral moments of $RR^\dagger$
for the NS sector of the massless free Dirac fermion.
\subsection{Charged moments}
\begin{figure}[t!]
    \centering
    \includegraphics[width=\textwidth]{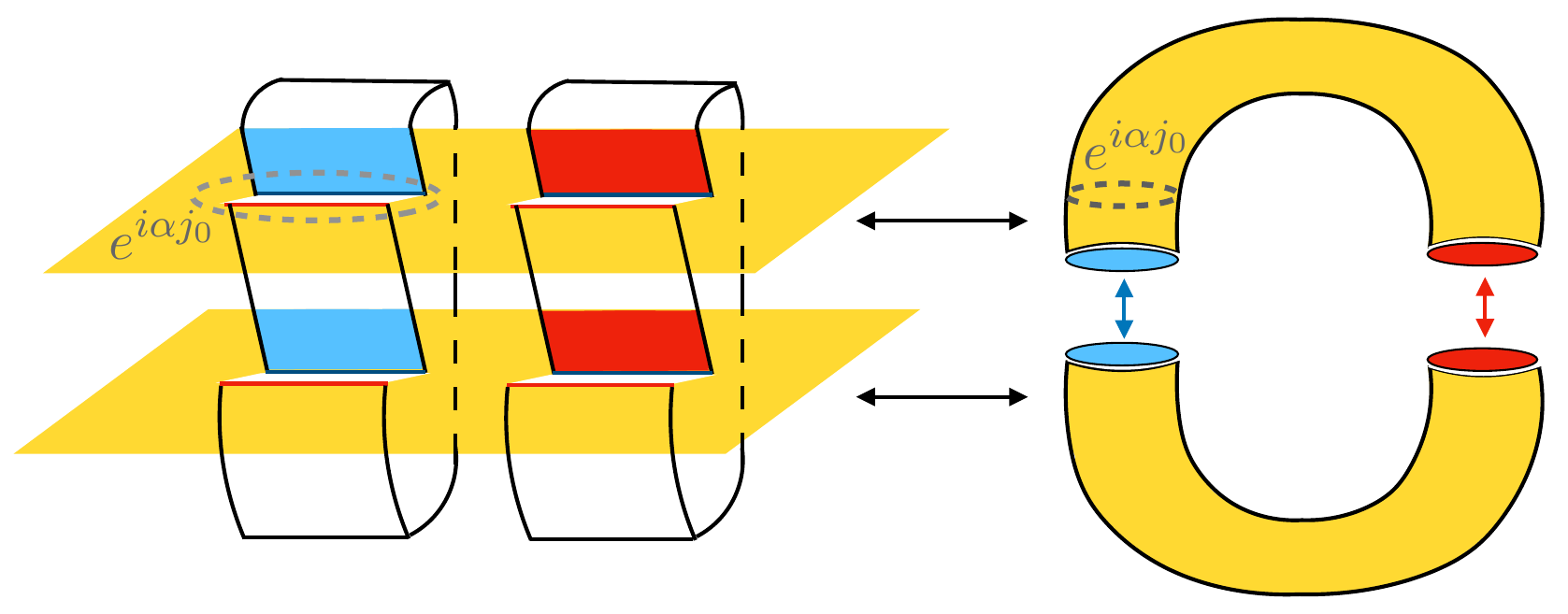}
    \caption{On the left, we represent the Riemann surface $\mathcal{M}_2$ arising in the 
    calculation of the charged moment $Z_1(\alpha)=\Tr(RR^\dagger e^{i\alpha \mathcal{Q}_A})$. The operator $e^{i\alpha \mathcal{Q}_A}$ corresponds to the insertion in the surface of a charged loop that encloses the branch cut corresponding to the interval $A$. Under the conformal transformation~\eqref{map}, each sheet of the surface $\mathcal{M}_{2}$ is mapped to a half of a flat torus, as we illustrate in the figure on the right. The boundary conditions that connect the two sheets correspond to gluing the two halves of the torus. The charged loop around the branch cut of the interval $A$ is mapped to one of the non-contractible cycles of the flat torus.}
    \label{fig:glue}
\end{figure}

Let us now calculate the two-interval charged moments $Z_n(\alpha)$ in the NS sector of the massless Dirac theory~\eqref{eq:lagrangianDirac}. Following the steps we took to derive the total CCNR negativity in Sec.~\ref{sub:total}, we start from the $n=1$ case. 
For this purpose, it is useful to explicitly express the charged moment $Z_1(\alpha)$ in terms of the reduced density matrix $\rho_{AB}$. In fact, as we prove in Appendix~\ref{app:proofs}, $Z_1(\alpha)$ can be rewritten as
\begin{equation}\label{eq:rewriting}
    Z_1(\alpha)=\mathrm{Tr} [RR^{\dagger}e^{i\alpha \mathcal{Q}_A}]=\mathrm{Tr}[\rho_{AB}e^{i\alpha Q_A}\rho_{AB}e^{-i\alpha Q_A}].
\end{equation}
The right hand side in Eq.~\eqref{eq:rewriting} can be interpreted as a path integral on the Riemann surface $\mathcal{M}_{2}$, depicted on left hand side of Fig.~\ref{fig:glue}, in which the operators $e^{i\alpha Q_A}$ and $e^{-i\alpha Q_A}$ are implemented as a two charged lines along each side of the branch cut corresponding to the interval $A$, joining its end-points $u_a$, $v_a$, and with opposite orientations.
Since $[Q_A+Q_B,\rho_{AB}]=0$ but $[Q_A,\rho_{AB}]\neq 0$, it is important to respect the order of the insertion of these two charged lines, which otherwise would simply fuse into the identity operator. In fact, the insertion of the two charged lines in Eq.~\eqref{eq:rewriting} creates a non-contractible charged loop around the branch cut of the interval $A$, as we show in Fig.~\ref{fig:glue}. When we map the Riemann surface $\mathcal{M}_2$ to the flat torus using Eq.~\eqref{map}, each sheet is sent to half of the torus, as we illustrate on the right side of Fig.~\ref{fig:glue}, where we also highlight the charged loop. After gluing together the two halves of the torus, which corresponds to connecting the two sheets of $\mathcal{M}_{2}$ as shown in the left picture, we find that $Z_1(\alpha)$ is equivalent to computing the partition function $Z(\tau,\alpha)$ on a flat torus of modulus $\tau$ with a closed charged loop around one of its non-contractible cycles.
If we take into account the Weyl anomaly of the map~\eqref{map} between the surface $\mathcal{M}_{2}$ and the flat torus, see Eq.~\eqref{eq:Ztau}, then we have
\begin{equation}\label{eq:charged_mom_n_1}
Z_1(\alpha)=c_1(\alpha) \frac{ Z(\tau, \alpha)}{[\ell_b\ell_a(u_b-u_a)(v_b-v_a)(v_b-u_a)(u_b-v_a)]^{c/12}},
\end{equation}
where $c_1(\alpha)$ is a non-universal constant that depends on the 
particular lattice model considered.

The partition function of the massless Dirac fermion~\eqref{eq:lagrangianDirac} on the flat torus with NS boundary conditions and a charged loop along the non-contractible cycle $a$ can be computed as~\cite{blumenhagen}
\begin{equation}
    Z(\tau, \alpha) =\left|\mathrm{Tr}_{\mathcal{H_{NS}}}[q^{L_0-\frac{c}{24}}e^{i\alpha j_0}]\right|^2, 
    \label{circa}
\end{equation}
where $j_0$ is the zero-mode of the conserved fermionic current $\psi\bar{\psi}$.
After bosonisation, we can write down the latter in terms of the bosonic field $\phi(z)$ as 
\begin{equation}
    j_0=\oint_a \frac{dz}{2\pi i} i\partial \phi(z),
    \label{loopeq}
\end{equation}
and $e^{i\alpha j_0}$ implements the charged loop on the flat torus. By recalling how the operators $L_0$ and $j_0$ act on all the states of the Hilbert space of the theory~\cite{blumenhagen}, we find
\begin{equation}
    \mathrm{Tr}_{\mathcal{H_{NS}}}[q^{L_0-\frac{c}{24}}e^{i\alpha j_0}]=\frac{1}{\eta(\tau)}\sum\limits_{n}q^{\frac{n^2}{2}}e^{i\alpha n}=\frac{\theta_3(\frac{\alpha}{2}|\tau)}{\eta(\tau)}.
\end{equation}
Inserting this result into Eq.~\eqref{eq:charged_mom_n_1}, we obtain 
\begin{equation}\label{eq:Z_1_alpha}
    Z_1(\alpha)=\frac{c_1(\alpha)}{[\ell_b\ell_a(u_b-u_a)(v_b-v_a)(v_b-u_a)(u_b-v_a)]^{c/12}}\left|\frac{\theta_3(\frac{\alpha}{2}|\tau)}{\eta(\tau)}\right|^2.
\end{equation}

The generalisation of Eq.~\eqref{eq:Z_1_alpha} to $n\neq 1$ can be obtained in a similar way as done in~\cite{yl-23} for the neutral moments and discussed in Sec.~\ref{sub:total} by re-scaling the torus modular parameter $\tau \to n \tau$ and the central charge $c\mapsto nc$, which yields
\begin{equation}\label{eq:final_charged}
    Z_n(\alpha)=\frac{c_1(\alpha)^n}{[\ell_b\ell_a(u_b-u_a)(v_b-v_a)(v_b-u_a)(u_b-v_a)]^{nc/12}}\left|\frac{\theta_3(\frac{\alpha}{2}|n\tau)}{\eta(n\tau)}\right|^2.
\end{equation}
Notice that generically the non-universal constant $c_1(\alpha)$ could depend on $\alpha$. In what follows, we will 
fix it by studying the limit in which the intervals $A$ and 
$B$ are far away.

\paragraph{Adjacent and far interval limits:} In the case in which the distance $d=|v_a-u_b|$ between the two intervals $A$ and $B$ is large, we have $ x\approx 16e^{i \pi\tau } \to 0 $ and  $\tau\rightarrow i\infty$. Therefore, the charged moments in Eq.~\eqref{eq:final_charged} behave as
\begin{equation}\label{eq:farlimit}
    \frac{Z_n(\alpha)}{Z_n(0)} \sim \frac{c_1(\alpha)^n}{c_1^n}\left( 1-4e^{2\pi in \tau}(1-\cos\alpha)\right).
\end{equation}
In this limit, the reduced density matrix $\rho_{AB}$ factorises as 
$\rho_{AB}\to \rho_A\otimes \rho_B$ and, consequently, the commutator $[\rho_{AB}, Q_A]\to 0$. Applying this in the identity of Eq.~\eqref{eq:rewriting}, we find that $Z_n(\alpha)\to Z_n(0)$ when $d\to\infty$. Taking this result into account in Eq.~\eqref{eq:farlimit}, we conclude that $c_1(\alpha)=c_1$. Therefore, for large subsystem lengths, the quotient $Z_n(\alpha)/Z_n(0)$ does not contain non-universal terms that depend on the lattice realisation and it is fully determined by conformal invariance. 

On the other hand, if the distance between the two intervals tends to zero, i.e. $d\to \epsilon$ (where $\epsilon\ll 1$ is an ultraviolet (UV) cutoff), 
then $x\sim 1-e^{-\frac{i\pi}{\tau}}\rightarrow 1$ and $\tau\to i 0^+$. In this regime, we find that
\begin{equation}\label{eq:adjacent}
    \frac{Z_n(\alpha)}{Z_n(0)}\sim e^{-\frac{\alpha^2}{2\pi |n\tau|}} \left(\frac{e^{\frac{\pi}{|n\tau|}}+2\cosh{\frac{\alpha}{|n\tau|}}}{2+e^{\frac{\pi}{|n\tau|}}}\right)^2\simeq \bigg(\frac{\epsilon(\ell_a+\ell_b)}{\ell_a\ell_b}\bigg)^{\frac{\alpha^2}{2n\pi^2}}.
\end{equation}
This result matches the one found in Ref.~\cite{bp-23} for two adjacent intervals and reported in Eq.~\eqref{eq:gilles}. Contrarily to the limit of far intervals, we stress that the term $\epsilon^{\alpha^2/(2\pi^2 n)}$ is not universal and it depends on the lattice realisation of the field theory that we consider.
As already mentioned, Eq.~\eqref{eq:gilles} can be obtained from the three-point 
correlator~\eqref{eq:vertex}. It is interesting to see how the latter arises from 
the partition function on $\mathcal{M}_{2n}$ with a non-contractible
charged loop. For simplicity, let us consider the case $n=1$ represented on the left hand side of Fig.~\ref{fig:glue}. In the limit of adjacent intervals, the branch points $v_a$ and $u_b$ coalesce, and the Riemann surface $\mathcal{M}_{2}$ is pinched. This pinching cuts the charged loop, which becomes a charged line with the two end-points at $v_a=u_b$, each one in a different sheet of $\mathcal{M}_2$.
The charged line can be implemented by a couple of vertex operators $\mathcal{V}_{\pm\alpha}(z)=e^{\pm i\frac{\alpha}{2\pi} \phi(z)}$, with conformal dimension $h_{\alpha}=\alpha^2/(8\pi^2)$, at its end-points. The correlator~\eqref{eq:corr_function_adjacent} corresponds to 
the (regularised) partition function without the loop after the pinching. If we plug the operator $(\mathcal{V}_{\alpha}\otimes \mathcal{V}_{-\alpha})(u_b)$ that describes the cut loop in it, then its fusion with the twist fields at the same point yields the composite fields $\mathcal{T}_{n,\alpha}\equiv \mathcal{T}_n\cdot \mathcal{V}_{\alpha}$ with dimension~\eqref{eq:dim_composite_fields}, see Ref.~\cite{goldstein}, and Eq.~\eqref{eq:vertex} directly follows.

\begin{figure}
    \centering
    \includegraphics[width=0.49\textwidth]{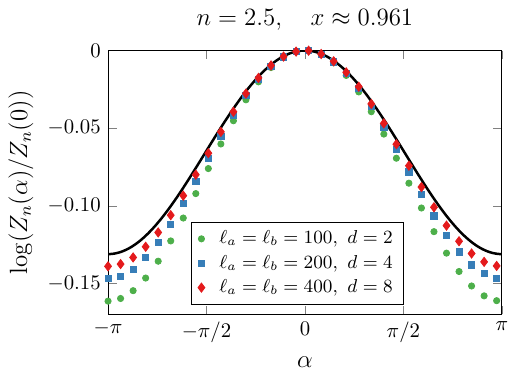}
    \includegraphics[width=0.49\textwidth]{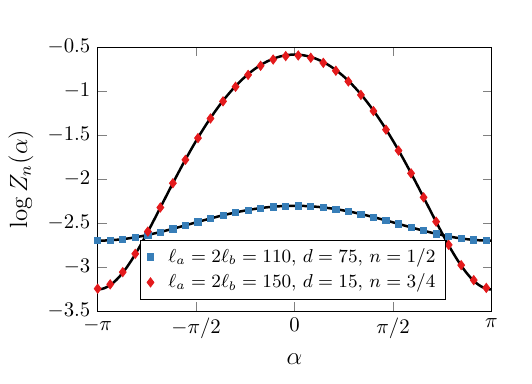}
    \caption{Left panel: Study of the convergence to the CFT prediction~\eqref{eq:ratio} of the quotient $Z_n(\alpha)/Z_n(0)$ in the tight-biding model when the cross-ratio $x$ is close to 1. We plot $Z_n(\alpha)/Z_n(0)$ versus $\alpha$ for different sets of $\ell_a$, $\ell_b$ and $d$ that give the same cross-ratio $x\approx 0.961$. The symbols are the exact numerical values for the ground state of tight-binding model, calculated as described in Appendix~\ref{app:numerics}, and the solid line is the CFT prediction~\eqref{eq:ratio}. Right panel: Plot of the realignment charged moments $\log Z_n(\alpha)$ as a function of $\alpha$ for two different values of $n$. The solid line is the CFT prediction of Eq.~\eqref{eq:final_charged} taking the non-universal constant $c_1(\alpha)$ determined in the main text for the tight-binding model. The symbols are the exact numerical value of $\log Z_n(\alpha)$ in the ground state of the tight binding model, calculate using the methods of Appendix~\ref{app:numerics}.}    
    \label{fig:convergenzatoCFT}
\end{figure}

\paragraph{Numerical checks:}
\begin{figure}[t]
\centering
\vspace{2mm}
{\includegraphics[width=0.485\textwidth]{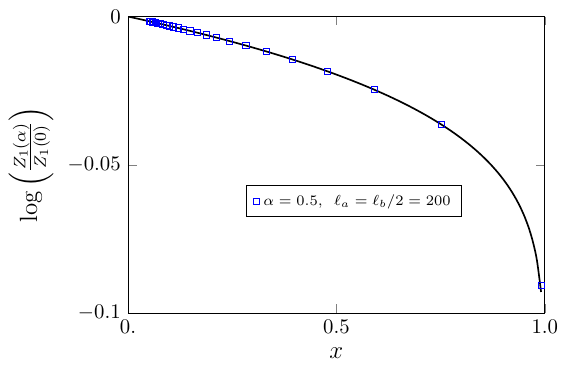}
    }
\hfill
{\includegraphics[width=0.495\textwidth]{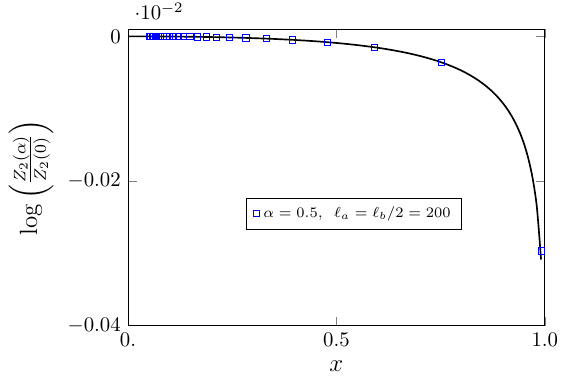}
    }
\hfill
{\includegraphics[width=0.49\textwidth]{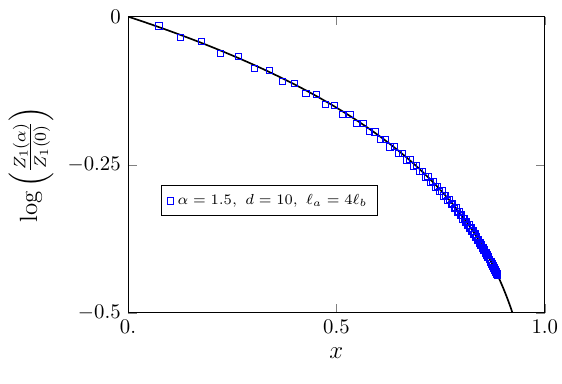}
    }
\hfill
{\includegraphics[width=0.49\textwidth]{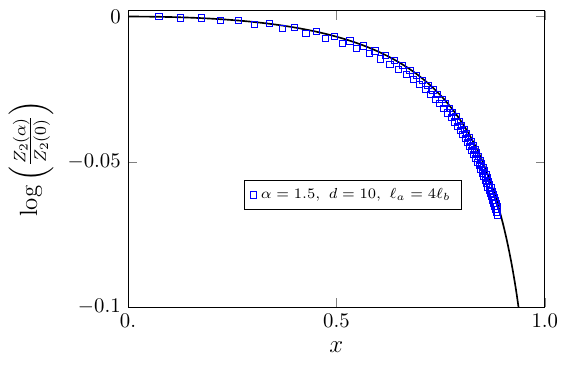} }
    \caption{Plot of $\log [Z_n(\alpha)/Z_n]$ as a function of the cross-ratio $x$ for different values of $\alpha$ and $n $. The symbols are the exact numerical value obtained for the ground state of the tight-binding model as described in Appendix \ref{app:numerics}. The cross-ratio $x$ is varied by keeping fixing $\ell_a=\ell_b/2=200$ and changing their separation $d$ (top panels) or fixing the distance $d=10$ and varying the subsystem size $\ell_a=4\ell_b$ (bottom panels). The solid black lines are the CFT analytical prediction in Eq.~\eqref{eq:ratio}.}
    \label{plots}
\end{figure}
The analytical prediction for the charged moments in Eq.~\eqref{eq:final_charged} can be benchmarked using the tight-binding model as microscopical realisation of the action in Eq.~\eqref{eq:lagrangianDirac} in the NS sector. We refer to Appendix~\ref{app:numerics} for the details about the numerical implementation of the problem.
In Fig.~\ref{plots}, we study $\log(Z_n(\alpha)/Z_n(0))$ as a function
of the cross-ratio $x$, defined in Eq.~\eqref{crossratio}, for different values of the subsystem lengths $\ell_a$, $\ell_b$, the distance $d$ between them, and the flux $\alpha$. In particular, the solid curves correspond to the analytic prediction of Eq.~\eqref{eq:final_charged},
\begin{equation}\label{eq:ratio}
   \frac{Z_n(\alpha)}{Z_n(0)}=\left|\frac{\theta_3(\frac{\alpha}{2}|n\tau)}{\theta_3(0|n\tau)}\right|^2,
\end{equation}
where we took $c_1(\alpha)=c_1$ as we concluded from the far interval limit, showing a perfect agreement with the 
exact numerical values (symbols) for the ground state of the tight-binding model. We remark that Eq.~\eqref{eq:ratio} 
is universal, and we do not have to fit any parameter to match the 
exact results from lattice computations, in agreement with what we 
expected. When $x\to 1$, i.e. the limit of adjacent intervals, the 
matching between the CFT expression~\eqref{eq:ratio} and the exact 
lattice results worsens since in this limit Eq.~\eqref{eq:ratio} diverges if we do not regularise it with a non-universal UV cutoff $\epsilon$ as we saw in Eq.~\eqref{eq:adjacent}, and one has to consider larger $\ell_a$, $\ell_b$ and $d$ to improve the agreement. We show this in the left panel of Fig.~\ref{fig:convergenzatoCFT}, where we choose a 
fix value of the cross-ratio $x$ close to 1, and we analyse 
$Z_n(\alpha)/Z_n(0)$ as a function of $\alpha$
when re-scaling the interval lengths $\ell_a$, $\ell_b$, and the distance $d$. 

Finally, in the right panel of Fig.~\ref{fig:convergenzatoCFT}, we directly bolster the analytical prediction in Eq.~\eqref{eq:final_charged} by plotting it as a function of $\alpha$ for two fixed cross-ratios $x$. For the tight-biding model, we can determine the non-universal factor $c_1(\alpha)$. We have concluded 
and checked numerically in Fig.~\ref{plots} that $c_1(\alpha)=c_1$. According 
to Eq.~\eqref{eq:rewriting}, $Z_1(0)=\Tr(\rho_{AB}^2)$ and, therefore, $c_1$ is 
the equal to the non-universal constant of the purity $\Tr(\rho_{AB}^2)$. The latter has been computed for the ground state of the tight-binding-model in e.g. Ref. \cite{ares14} applying the asymptotic properties of Toeplitz determinants~\cite{jk-04} and reads $c_1=e^{4\Upsilon_2}$, where $\Upsilon_2\approx -0.404049$. In the right panel of Fig.~\ref{fig:convergenzatoCFT}, the solid curves correspond to Eq.~\eqref{eq:final_charged} using it as non-universal constant, while the symbols are the exact numerical value of $Z_n(\alpha)$ in the ground state of the tight-binding model. We obtain an excellent agreement.

\subsection{Symmetry-resolved CCNR negativity}
The final step to get the symmetry-resolved CCNR negativity in Eq.~\eqref{eq:def_SRCCNR} is to take the Fourier transform~\eqref{eq:ft} of the charged moments $Z_n(\alpha)$ in Eq.~\eqref{eq:final_charged}. Let us study the adjacent and disjoint cases separately. 

In the former situation, the result has been derived in Ref.~\cite{bp-23} and we have reported it in Eq.~\eqref{eq:gillesadj}. Therefore, we can start by a numerical benchmark of this prediction in the tight-binding model, using again the techniques described in Appendix~\ref{app:numerics}. In the left panel of Fig. \ref{fig:adjacent}, we compute numerically $\mathcal{E}(q)-\mathcal{E}$ for the ground state of this model and we plot it as a function of $y=\log \frac{\ell_a\ell_b}{\ell_a+\ell_b}$: it shows that, 
increasing $\ell_a$, $\mathcal{E}(q)$ converges to the right hand side of Eq.~\eqref{eq:gillesadj}, since $\mathcal{E}(q)-\mathcal{E}+1/4 \log (8y/\pi)$ approaches zero, especially for smaller values of the charge $q$. However, it is clear that, to have a more refined comparison with the lattice computations in the tight-binding model, we have to take into account the non-universal factor that appears in this limit and its $\alpha$-dependence (see Eq.~\eqref{eq:adjacent}). For this purpose, we can rewrite Eq.~\eqref{eq:adjacent} as
\begin{equation}  Z_n(\alpha)=c_n(\alpha)\bigg(\frac{\ell_a+\ell_b}{\ell_a\ell_b}\bigg)^{\frac{\alpha^2}{2n\pi^2}}\left(\ell_a\ell_b\right)^{-\frac{1}{6}(n-\frac{1}{n})}(\ell_a+\ell_b)^{-\frac{1}{12}\left(n+\frac{2}{n}\right)},
    \label{malsana}
\end{equation}
where $c_n(\alpha)=c_n\epsilon^{\frac{\alpha^2}{2n\pi^2}}$ is the non-universal term. From the definition of $c_n(\alpha)$, it directly follows $c_n(0)=c_n$. Applying the approach of Appendix~\ref{app:numerics}, we can
study numerically in the tight-binding model the dependence of $c_n(\alpha)$ on $\alpha$ for a given value of $n$ and we find that it can be well approximated as $c_n(\alpha)\simeq c_ne^{b_n\alpha^2}$. 
Using Eq.~\eqref{malsana} with the values of $c_n$ and $b_n$ extracted from a fit of the exact data to this function, we perform numerically the Fourier transform~\eqref{eq:ft} and we plot Eq.~\eqref{eq:SRCCNR_fourier} in Fig.~\ref{fig:adjacent} (solid line). We also report the result obtained by neglecting the $\alpha$-dependent factor in $c_n(\alpha)$ (dashed lines), showing how the agreement with the exact numerical points (symbols) improves if we take it into account. 
Both for $n=1/2$ and $1/3$, $\mathcal{E}_n(q)$ is a positive quantity, but we have noticed that for larger values of $n$, $\mathcal{E}_n(q)$ can take negative values.

If the intervals $A$ and $B$ are disjoint, we can perform the Fourier transform~\eqref{eq:ft} of Eq.~\eqref{eq:final_charged} analytically and we find that 
\begin{equation}
    \frac{\mathcal{Z}_n(q)}{\mathcal{Z}^n_1(q)}=\begin{cases}
        &\frac{|\eta(\tau)|^{2n}}{|\eta(n\tau)|^2}\frac{\theta_3(0|2n\tau)}{\theta_3(0|2\tau)^n}, \qquad q\,\mathrm{ even},\\
        & \frac{|\eta(\tau)|^{2n}}{|\eta(n\tau)|^2}\frac{\theta_2(0|2n\tau)}{\theta_2(0|2\tau)^n}, \qquad q\,\mathrm{ odd}.
    \end{cases}
\end{equation}
Therefore, applying Eq.~\eqref{eq:SRCCNR_fourier}, the symmetry-resolved R\`enyi CCNR negativity reads 
\begin{equation}\label{eq:final_CCNRq}
    \mathcal{E}_n(q)=\log\frac{|\eta(\tau)|^{2n}}{|\eta(n\tau)|^2}+n\log \mathrm{Tr}(\rho_{AB}^2)+\begin{cases}
    &\log\frac{\theta_3(0|2n\tau)}{\theta_3(0|2\tau)^n} \qquad q\,\mathrm{ even}\\
        & \log\frac{\theta_2(0|2n\tau)}{\theta_2(0|2\tau)^n} \qquad q\,\mathrm{ odd},
    \end{cases}
\end{equation}
where $\mathrm{Tr}(\rho^2_{AB})=c_1 2^{2/3}(\ell_a\ell_b(1-x))^{-1/4}$~\cite{ares14, cfh-05}. This expression for the purity of $\rho_{AB}$ can be deduced taking into account 
that, according to Eq.~\eqref{eq:rewriting}, $Z_1(0)=\Tr(\rho_{AB}^2)$ and $Z_1(0)$ is given in our case by~\eqref{eq:final_charged}; note that here we have used the identities $x=(\theta_2(0|\tau)/\theta_3(0|\tau))^4$ and $2\eta(\tau)^3=\theta_2(0|\tau)\theta_3(0|\tau)\theta_4(0|\tau)$. 

\begin{figure}
    \centering
    \includegraphics[width=0.49\textwidth]{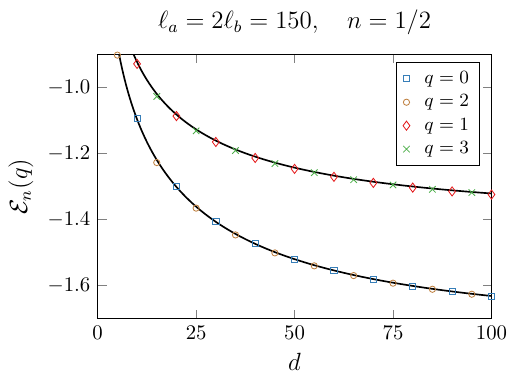}
    \includegraphics[width=0.49\textwidth]{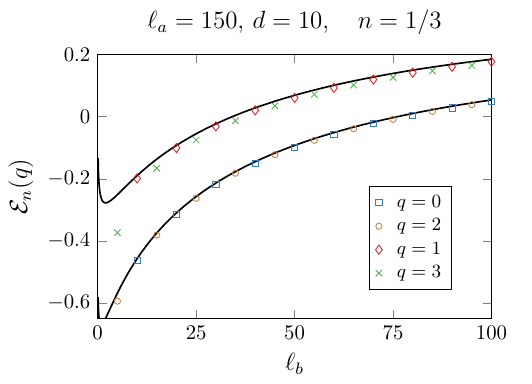}
    \caption{Symmetry-resolved CCNR negativity $\mathcal{E}_n(q)$ for two disjoint intervals as a function of the distance $d$ for R\'enyi index $n=1/2$ (left panel) and changing the lenght $\ell_b$ of interval $B$ taking R\'enyi index $n=1/3$ (right panel). In both cases, the symbols correspond to the exact values calculated for the ground state of the tight-binding model as we explain in Appendix~\ref{app:numerics} and the solid lines represent the 
    CFT analytic prediction in Eq.~\eqref{eq:final_CCNRq}.}
    \label{fig:disjoint}
\end{figure}

The result of Eq.~\eqref{eq:final_CCNRq} is quite remarkable since it predicts that $\mathcal{E}_n(q)$ is exactly equally distributed among the different charge sectors up to their parity. This is not the first time that the symmetry resolution of the entanglement shows exact equipartition up to the parity of the charge: for instance, this behaviour has been observed in Ref.~\cite{ccdgm-20}, in the gapped phase of the XXZ spin chain, and in Ref.~\cite{hfsr-23}, studying the Su-Schrieffer-Heeger model in the presence of topological defects. Both examples consider the symmetry resolution of the total entanglement entropy in systems away from criticality, while here we are analysing a critical system. Therefore, the origin of the dependence on the parity of the charge is a purely universal feature of the CCNR negativity, in the sense that it is univocally derived from the CFT prediction for the charged moments in Eq. \eqref{eq:ratio} and it is not related to some specific lattice computations, as it instead happens for the symmetry-resolved entanglement entropy \cite{ccdgm-20,hfsr-23}. In Fig.~\ref{fig:disjoint}, we benchmark the analytical prediction of Eq.~\eqref{eq:final_CCNRq} with exact numerical computations in the ground state of the tight-binding mode, finding a good agreement.

\begin{figure}[t]
    \centering
    \includegraphics[width=0.49\textwidth]{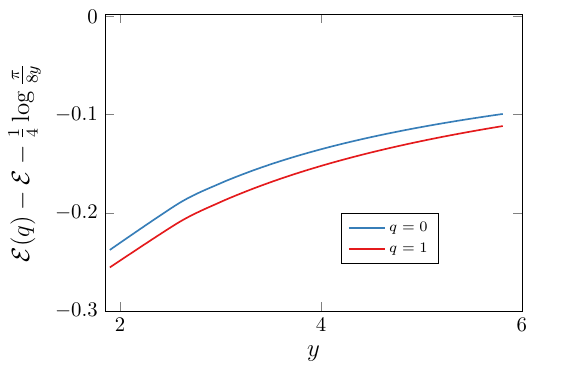}
    \includegraphics[width=0.49\textwidth]{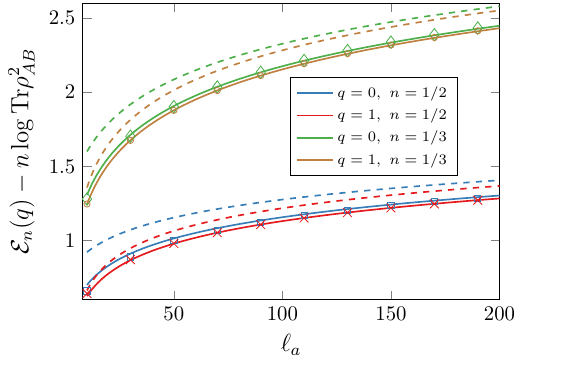}
    \caption{Left panel: Check of the prediction in Eq.~\eqref{eq:gillesadj} for the symmetry-resolved CCNR negativity
    of two adjacent intervals. We calculate numerically $\mathcal{E}(q)-\mathcal{E}$ taking $\ell_b=2\ell_a$ for the ground state of the tight-binding model 
    and we plot the result
    as a function of $y=\log \frac{\ell_a\ell_b}{\ell_a+\ell_b}$. The plot shows that increasing $\ell_a$, $\mathcal{E}(q)$ converges to the right hand side of Eq.~\eqref{eq:gillesadj}, which has been obtained neglecting all the non-universal contributions.
    Right panel: Symmetry-resolved CCNR R\'enyi negativity $\mathcal{E}_n(q)$ for two adjacent intervals of lengths $\ell_b=2\ell_a$ varying $\ell_a$. The (dashed) solid lines represent Eq.~\eqref{eq:SRCCNR_fourier} obtained from the expression of the charged moments in Eq.~\eqref{malsana} for adjacent intervals (without) taking into account the non-universal factor $c_n(\alpha)$, which we determine as described in the main text. The symbols are the exact numerical values for the ground state of the tight-binding model. }
    \label{fig:adjacent}
\end{figure}

\section{Conclusions}\label{sec:conclusions}

In this manuscript, we have investigated the symmetry resolution of the CCNR negativity in the presence of a global $U(1)$ symmetry. This work was initiated in Ref.~\cite{bp-23} for two adjacent intervals  $A$ and $B$ in the ground state of the massless Dirac fermion field theory and here we have extended it to a generic configuration of the two regions. For this purpose, we have leveraged the charged moments of $RR^{\dagger}$, where $R$ is the two-interval realignment matrix. In Sec.~\ref{sec:new}, we have shown that they can be computed as a partition function on a genus-1 surface (a torus) with the insertion of a charged loop along one of its non-contractible cycles for any value of the replica index $n$. This makes the calculation easier with respect to other quantities that are relevant to probe entanglement in mixed states, such as the negativity defined in terms of the partial transpose, which involves the evaluation of partition functions on higher-genus surfaces. A very interesting result that we have obtained is that, when $A$ and $B$ are distant enough, the symmetry-resolved CCNR R\'enyi negativities, $\mathcal{E}_n(q)$, depend on the parity sector of the charge $q$ while, if they are adjacent, $\mathcal{E}_n(q)$ is independent of the charge and satisfies an exact equipartition. Despite this result, we have observed that, when $A$ and $B$ are disjoint intervals, $\mathcal{E}_n(q)$ can be negative for different values of $n$, including the replica limit $n\to 1/2$. This can be related to the fact the CCNR negativity is not an entanglement measure, as it was proven in~\cite{bp-23}.

We have compared all our analytical predictions with exact lattice computations using the tight-binding model, showing how they capture the universal behaviour of the charged moments of the realignment matrix and the symmetry-resolved CCNR negativity. However, in order to have a good matching between the CFT predictions and the exact values on the lattice, one has to take into account the non-universal terms specific to the tight-binding model. We were able to find their analytical expression for disjoint intervals, while in the adjacent geometry we could only estimate them through a numerical fit. This connects to one of the first future possible directions that one could pursue, that is, the computation of the symmetry-resolved CCNR negativity on a lattice model in which all the non-universal sub-leading terms can be exactly worked out (as done in Ref.~\cite{riccarda} for the symmetry-resolved entanglement in the tight-binding model).

There are many other possibilities to further investigate the realignment matrix and the CCNR negativities. For example, given the universal behaviour found in Ref.~\cite{yl-23} for the moments of the realignment matrix in CFTs, it would be interesting to study its spectrum and check whether it still preserves the universal character as happens for the reduced density matrix~\cite{cl-08} and its partial transpose~\cite{rac-16}. It is also natural to extend the symmetry-resolution of the CCNR negativity to a free massless compact boson, which also enjoys a global $U(1)$ symmetry, and study how the dependence on the compactification radius modifies our predictions. Another appealing direction is the generalisation of the CCNR negativity to more than two intervals and to investigate whether it is also possible to relate it to Riemann surfaces with the topology of the torus.  

\subsection*{Acknowledments}
We thank Clement Berthiere, Jerome Dubail, Michele Fossati, Gilles Parez and Federico Rottoli for useful discussions. PC and FA acknowledge support from ERC under Consolidator Grant number 771536 (NEMO). SM thanks the support from the Caltech Institute for Quantum Information and Matter and the Walter Burke Institute for Theoretical Physics at Caltech.

\appendix
\section*{Appendices}

\section{Proof of the identities in Eqs.~\eqref{eq:ReqOE} and~\eqref{eq:rewriting}}\label{app:proofs}

In this Appendix, we prove the identities in Eqs.~\eqref{eq:ReqOE} and
\eqref{eq:rewriting}, which we apply in several crucial points of the main text. Let $\{\ket{{\rm a}}\}$ and $\{\ket{{\rm b}}\}$ be bases of $\mathcal{H}_A$ and $\mathcal{H}_B$ respectively, then the realignment matrix can be written according to its definition~\eqref{eq:def_realignment} in the form
\begin{equation}\label{eq:realignment_op}
R=\sum_{\substack{{\rm a, a'} \\ {\rm b, b'}}}\bra{{\rm ab}}\rho_{AB}\ket{{\rm a'b'}}\ket{{\rm aa'}}\bra{{\rm bb'}}.
\end{equation}

Let us first consider the identity~\eqref{eq:ReqOE}. If we calculate using~\eqref{eq:realignment_op} the product $RR^\dagger$, we find
\begin{equation}\label{eq:RRdagger}
RR^\dagger=\sum_{\substack{{\rm a, a'}\\ \tilde{{\rm a}}, \tilde{{\rm a}}'}}\sum_{{\rm b}, {\rm b'}}\bra{{\rm ab}}\rho_{AB}\ket{{\rm a' b'}}
\bra{\tilde{{\rm a}}'{\rm b}'}\rho_{AB}\ket{\tilde{\rm a}{\rm b}}\ket{{\rm aa'}}\bra{\tilde{{\rm a}}\tilde{{\rm a}}'},
\end{equation}
where we have taken into account that $\rho_{AB}$ is Hermitian and, therefore, $\bra{\tilde{{\rm a}} {\rm b}}\rho_{AB}\ket{\tilde{{\rm a}}'{\rm b}'}^*=\bra{\tilde{{\rm a}}'{\rm b}'}\rho_{AB}\ket{\tilde{{\rm a}} {\rm b}}$. On the other hand, if we take the vector $\ket{\rho_{AB}}$ associated to $\rho_{AB}$ by the Choi-Jamiolkowski isomorphism~\eqref{eq:vect}, we have 
\begin{equation}
\ket{\rho_{AB}}\bra{\rho_{AB}}=\frac{1}{\Tr(\rho_{AB}^2)}
\sum_{\substack{{\rm a, a'}\\ {\rm b, b'}}}
\sum_{\substack{\tilde{{\rm a}}, \tilde{{\rm a}}'\\ \tilde{{\rm b}}, \tilde{{\rm b}}'}}\bra{{\rm ab}}\rho_{AB}\ket{{\rm a' b'}}
\bra{\tilde{{\rm a}}'\tilde{{\rm b}}'}\rho_{AB}\ket{\tilde{\rm a}\tilde{{\rm b}}}\ket{{\rm ab}}\ket{{\rm a' b'}}\bra{\tilde{{\rm a}}\tilde{{\rm b}}}\bra{\tilde{{\rm a}}'\tilde{{\rm b}}'}.
\end{equation}
Taking in this expression the partial trace in the subspace $\mathcal{H}_B\otimes \mathcal{H}_B$,  $\sigma_{AA'}=\Tr_{BB'}(\ket{\rho_{AB}}\bra{\rho_{AB}})$, we obtain
\begin{equation}
  \sigma_{AA'}=\frac{1}{\Tr(\rho_{AB}^2)}
    \sum_{\substack{{\rm a, a'}\\ \tilde{{\rm a}}, \tilde{{\rm a}}'}}\sum_{{\rm b}, {\rm b'}}\bra{{\rm ab}}\rho_{AB}\ket{{\rm a' b'}}
\bra{\tilde{{\rm a}}'{\rm b}'}\rho_{AB}\ket{\tilde{\rm a}{\rm b}}\ket{{\rm aa'}}\bra{\tilde{{\rm a}}\tilde{{\rm a}}'}.
\end{equation}
Comparing this result with Eq.~\eqref{eq:RRdagger}, the identity~\eqref{eq:ReqOE} follows.

Regarding Eq.~\eqref{eq:rewriting}, let us assume that the elements of the basis $\{\ket{{\rm a}}\}$ of $\mathcal{H}_{A}$ are eigenvectors of the 
charge operator $Q_A$ in the interval $A$ with eigenvalue $q_a$, $Q_A\ket{{\rm a}}=q_{{\rm a}}\ket{{\rm a}}$. If we apply Eq.~\eqref{eq:RRdagger} in the charged moment $Z_1(\alpha)=\Tr(RR^\dagger e^{i\alpha \mathcal{Q}_A})$, and we recall the definition~\eqref{eq:superQ} of the super-charge operator $\mathcal{Q}_A$, we have
\begin{equation}
\Tr(RR^{\dagger}e^{i\alpha \mathcal{Q}_A})=\sum_{\substack{{\rm a, a'}\\ {\rm b, b'}}}\bra{{\rm ab}}\rho_{AB}\ket{{\rm a' b'}}
\bra{{\rm a}'{\rm b}'}\rho_{AB}\ket{{\rm a} {\rm b}}e^{i\alpha(q_{\rm a}-q_{{\rm a}'})},
\end{equation}
which can be rewritten as
\begin{equation}
\Tr(RR^\dagger e^{i\alpha \mathcal{Q}_A})=
\sum_{\substack{{\rm a}, {\rm a}' \\ {\rm b}, {\rm b}'}}
\bra{{\rm ab}}\rho_{AB}e^{i\alpha Q_A}\ket{{\rm a' b'}}
\bra{{\rm a}'{\rm b}'}\rho_{AB} e^{-i\alpha Q_A}\ket{{\rm a} {\rm b}}.
\end{equation}
This last expression is equal to $\Tr(\rho_A e^{i\alpha Q_A}\rho_A e^{-i \alpha Q_A})$ and, therefore, we obtain Eq.~\eqref{eq:rewriting}.

\section{Numerical methods}\label{app:numerics}
As a microscopical realisation of the free Dirac massless field theory with NS-NS boundary conditions, we consider the tight-binding model described by the Hamiltonian
\begin{equation}
    H=-\sum_{i} c_{i}^{\dagger}c_{i+1}+ \text{h.c.},
    \label{tight}
\end{equation}
where $c_{i},c_i^{\dagger}$ are fermionic operators satisfying the anticommutation relations $\{c_{i},c_i^{\dagger}\}=\delta_{ij}$.

By performing a Fourier transformation to momentum space $d_k=\sum_{j\in\mathbb{Z}} e^{-ikj} c_j$, Eq.~\eqref{tight} simplifies as
\begin{equation}\label{tight_k}
    H=-\sum_k \cos k \left(d^{\dagger}_kd_k-\frac{1}{2}\right),
\end{equation}
where $-\cos k$ is the single-particle dispersion relation. The ground state of this model, $\ket{\Omega}$, is the one that is annihilated by all the operators $d_k$, i.e. $d_k\ket{\Omega}=0$ for all $k$ such that $\cos (k)>0$.

The reduced density matrix for a subsystem $A\cup B$ can be diagonalised and put in the form $\rho_{AB} \propto e^{-\sum_k \lambda_kd^{\dagger}_k d_k}$, where $e^{-\lambda_k}=\frac{n_k}{1-n_k}$ with $n_k$ the occupation number of the $k$-th orbital. It can also be expressed as
\begin{equation} \rho_{AB}=\bigotimes_{k=1}^{\ell_a+\ell_b}[ (1-n_k) \ket{0_k}\bra{0_k}+n_k \ket{1_k}\bra{1_k}]
    \label{rhopeschel}
\end{equation}
where $\ket{1_k}=d_k^{\dagger}\ket{0_k}$.

After the vectorisation process in Eq.~\eqref{eq:vect} we can rewrite it 
\begin{equation}
\begin{split}
    \ket{\rho_{AB}} =\bigotimes_{k=1}^{\ell_a+\ell_b}\frac{[(1-n_k)\ket{0_k}\ket{\Tilde{0}_k}+n_k \ket{1_k} \ket{\Tilde{1}_{k}}]}{\sqrt{n_k^2+(1-n_k)^2}}=\bigotimes_{k=1}^{\ell_a+\ell_b}\frac{[(1-n_k)+n_k d_k^{\dagger} \Tilde{d}_{k}^{\,\dagger}]\ket{0_k}\ket{\Tilde{0}_{k}}}{\sqrt{n_k^2+(1-n_k)^2}},
    \label{shape}
    \end{split}
\end{equation}
where we use the notation $\Tilde{d}_k$ for operators acting on the Hilbert spaces $\mathcal{H}_{A'}$ and $\mathcal{H}_{B'}$. In momentum space, the correlation matrix of the state $\ket{\rho_{AB}}$ is~\cite{rath-23,zaletel-22}
\begin{equation}\label{eq:Ckk}
    C=\bigoplus_{k=1}^{\ell_a+\ell_b} C_{kk},
\end{equation}
where
\begin{equation} \label{eq:correl_mtx}
\begin{split}
C_{kk'}=\bra{\rho_{AB}} \begin{pmatrix}
d_k^{\dagger}\\
\tilde{d}_k
\end{pmatrix}
\begin{pmatrix}
d_{k'} \, \tilde{d}_{k'}^{\dagger}
\end{pmatrix}\ket{\rho_{AB}}=\frac{\delta_{kk'}}{n^2_k+(1-n_k)^2}\begin{pmatrix}
& n_k^2 &n_k(1-n_k) \\
&n_k(1-n_k)  & (1-n_k)^2 \\
\end{pmatrix}.
\end{split}
\end{equation}
The supercharge operator can be written in the basis of $d_k, \tilde{d}_k$'s and it takes the form 
\begin{equation}\label{eq:charge2}
\mathcal{Q}=(\sum_k d^{\dagger}_k d_k)\otimes \mathbbm{1}-\mathbbm{1}\otimes (\sum_k\tilde{d}^{\dagger}_k \tilde{d}_k)^T. 
\end{equation}
As we have discussed in Sec.~\ref{subsec:SR}, we are interested in the charged moments $Z_n(\alpha)=\mathrm{Tr}[(RR^{\dagger})^ne^{i\alpha\mathcal{Q}_A}]$. This can be traced back to the charged moments of the super reduced density matrix $\sigma_{AA'}$ using Eq.~\eqref{eq:ReqOE} and it can be written as $Z_n(\alpha)=[\mathrm{Tr}(\rho_{AB}^2)]^n\mathrm{Tr}[(\sigma_{AA'})^ne^{i\alpha\mathcal{Q}_A}]$.
To evaluate $\mathrm{Tr}[(\sigma_{AA'})^ne^{i\alpha\mathcal{Q}_A}]$, we start from the correlation matrix in Eq.~\eqref{eq:Ckk}, we do a Fourier transform to the spatial basis and we restrict it to the subsystem $A$. This restricted matrix has $2\ell_a$ eigenvalues, $\xi_i$, such that the charged moments read~\cite{rath-23,peschel2}
\begin{equation}\label{eq:lattice}
\mathrm{Tr}[\sigma_{AA'}^n e^{i\alpha\mathcal{Q}_A}]=e^{-i\alpha (\ell_A)}\prod_{a=1}^{2\ell_A}(\xi_a^{n}e^{i\alpha}+(1-\xi_a)^{n}).
\end{equation}

Putting everything together, we can compute the charged moments of the realignment as
\begin{equation}
    Z_n(\alpha)=e^{-i\alpha \ell_a} \bigg(\prod_{i=1}^{\ell_a+\ell_b} \Big[(1-n_k)^2+n_k^2  \Big] \bigg)^n \prod_{i=1}^{2\ell_a} \Big[(1-\xi_i)^n+e^{i\alpha}\xi_i^n  \Big],
    \label{numerics}
\end{equation}
and using Eq.~\eqref{eq:ft}, $Z_n(\alpha)$ is the starting point to study the symmetry resolution of the CCNR negativity.

\end{document}